\documentclass[twocolumn,showpacs,aps,prl,superscriptaddress,floatfix]{revtex4}

\usepackage{graphicx}
\usepackage{dcolumn}
\usepackage{amsmath}
\usepackage{epsfig}

\input pubboard/babarsym

\def\KKKz      {\ensuremath{K^+ K^- K^0}\xspace}
\def\KKKs      {\ensuremath{K^+ K^- \KS}\xspace}
\def\KKK      {\ensuremath{\Kp \Km \Kp}\xspace}
\def\KKKspm      {\ensuremath{ \Bz_{\scriptscriptstyle(+-)} }\xspace}
\def\KKKszz      {\ensuremath{ \Bz_{\scriptscriptstyle(00)} }\xspace}
\def\KKKll      {\ensuremath{ \Bz_{\scriptscriptstyle (L)} }\xspace}
\def\phiKz     {\ensuremath{\phi K^0}\xspace}
\def\mKK       {\ensuremath{m_{\Kp\Km}}\xspace}
\def\betaeff   {\ensuremath{\beta_{\mathit{eff}}}\xspace}
\def\betaeffr   {\ensuremath{\beta_{\mathit{eff},r}}\xspace}
\def\Acp   {\ensuremath{{A}_{\CP}}\xspace}
\def\Acpr   {\ensuremath{{A}_{\CP,r}}\xspace}
\def\cosH      {\ensuremath{\cos \theta_H}\xspace}
\def\BCP      {\ensuremath{\B_{\CP}}} 
\def\Btag      {\ensuremath{\B_{\mathrm{tag}}}} 
\def\qtag      {\ensuremath{q_{\mathrm{tag}}}} 
\def\Dsm      {\ensuremath{D^-_s}\xspace}
\def\fz      {\ensuremath{f_0}\xspace}
\def\DP      {DP\xspace}

\newcommand{\optbar}[1]{\shortstack{{\tiny(\rule[.4ex]{1em}{.1mm})}\\[-.7ex]$#1$}}

\def\AorAbar    {\kern 0.18em\optbar{\kern -0.18em {\cal A}}{}\xspace}
\def\BzorBzbar    {\kern 0.18em\optbar{\kern -0.18em B^0}{}\xspace}

\def\nbb {383}
\def\reflectionsigma{\ensuremath{4.5\sigma}\xspace} 
\def\CPsigma{\ensuremath{4.8\sigma}\xspace}   
\def\wholeDPbeta{\ensuremath{0.352 \pm 0.076 \pm 0.026}}
\def\wholeDPacp{\ensuremath{-0.015 \pm 0.077 \pm 0.053}}

\newcommand{\BABARPubYear}    {07}
\newcommand{\BABARPubNumber}  {029}

\newcommand{\SLACPubNumber} {12625}
\newcommand{\LANLNumber} {0706.3885}

\def\figurebox#1#2#3{%
    \def\arg{#3}%
    \ifx\arg\empyt
    {\hfill\vbox{\hsize#2\hrule\hbox to #2{\vrule\hfill\vbox to #1{\hsize#2\vfill}\vrule}\hrule}\hfill}%
    \else
    {\hfill\epsfbox{#3}\hfill}%
    \fi}

\begin{document}

\preprint{\babar-PUB-\BABARPubYear/\BABARPubNumber} 
\preprint{SLAC-PUB-\SLACPubNumber} 

\begin{flushleft}
\babar-PUB-\BABARPubYear/\BABARPubNumber\\
SLAC-PUB-\SLACPubNumber\\
arXiv:\LANLNumber\ [hep-ex]\\[10mm]
\end{flushleft}

\title{
{\large \bf Measurements of {\boldmath$\CP$}-Violating Asymmetries in
  the Decay {\boldmath $\Bz\to\Kp\Km\Kz$}} 
}

%
\author{B.~Aubert}
\author{M.~Bona}
\author{D.~Boutigny}
\author{Y.~Karyotakis}
\author{J.~P.~Lees}
\author{V.~Poireau}
\author{X.~Prudent}
\author{V.~Tisserand}
\author{A.~Zghiche}
\affiliation{Laboratoire de Physique des Particules, IN2P3/CNRS et Universit\'e de Savoie, F-74941 Annecy-Le-Vieux, France }
\author{J.~Garra~Tico}
\author{E.~Grauges}
\affiliation{Universitat de Barcelona, Facultat de Fisica, Departament ECM, E-08028 Barcelona, Spain }
% ...
\collaboration{The \babar\ Collaboration}
\noaffiliation

\date{\today}

\begin{abstract}
  We analyze the decay $\Bz
  \to \KKKz$ using \nbb\ million $\BB$ events collected by
  the \babar\ detector at SLAC to extract \CP violation parameter values
over the Dalitz plot. Combining all \KKKz events, 
  we find $\Acp = \wholeDPacp$ and $\betaeff = \wholeDPbeta\,\rad$,
  corresponding to a \CP violation significance of \CPsigma.
  A second solution near
  $\pi/2 - \betaeff$ is disfavored with a significance of
  \reflectionsigma. We also report \Acp and \betaeff separately for decays to
$\phi(1020)\Kz$, $f_0(980)\Kz$, and \KKKz with $\mKK>1.1 \gevcc$.
\end{abstract}


\maketitle

  In the Standard Model (SM), 
  the phase in the
  Cabibbo-Kobayashi-Maskawa (CKM) quark-mixing matrix~\cite{CKM} is the
  sole source of \CP violation in the quark sector. 
  Due to interference between decays with and without mixing, 
  this phase yields observable
  time-dependent \CP asymmetries in \Bz meson decays. 
  In particular, significant \CP asymmetries in $\b \to \s
  \sbar \s$ decays, such as $\Bz \to \KKKz$~\cite{conjugate}, are expected~\cite{Chen:2006nk,Aubert:2004zt}. 
Deviations from the predicted \CP asymmetry behavior for $\Bz
  \to \KKKz$ are expected to depend weakly on Dalitz plot (DP) 
  position~\cite{Beneke:2005pu,Buchalla:2005us}. 
  Since the $\b \to \s\sbar\s$
  amplitude is dominated by loop contributions, 
  heavy virtual particles beyond the SM
  might contribute significantly~\cite{Buchalla:2005us,newphysics}.
  This sensitivity motivates
  measurements of \CP asymmetries in multiple $\b \to \s \sbar \s$
  decays~\cite{Chen:2006nk,Aubert:2005ja,Aubert:2006wv,Chen:2005dr}.

  Previous measurements of \CP asymmetries in $\Bz \to \KKKz$
  have been performed separately 
  for events with $\Kp\Km$ invariant mass ($\mKK$) in the
  $\phi$ mass~\cite{assume1020} region,
  and for events excluding the $\phi$ region, 
  neglecting interference effects among intermediate
  states~\cite{Chen:2006nk,Chen:2005dr,Aubert:2005ja}.  
  In this Letter we describe a
  time-dependent \DP analysis of $\Bz \to \KKKz$
  decay from which we extract the values of the
  \CP violation parameters \Acp and $\betaeff$ by
  taking into account the complex amplitudes describing the entire \Bz and
  \Bzb Dalitz plots. 
We first extract the values of the parameters of the amplitude model,
and measure the
average \CP asymmetry in $\Bz \to \KKKz$ decay over the entire \DP.
Using this model, we then measure the \CP asymmetries for
the $\phi\Kz$ and $\fz\Kz$ decay channels, from a
``low-mass'' analysis of events with $\mKK < 1.1
\gevcc$. Finally, we perform a ``high-mass'' analysis to determine the
average \CP asymmetry for events with $\mKK>1.1\gevcc$.

  The data sample for this analysis was collected with the \babar\
  detector~\cite{ref:babar} at the \pep2\ asymmetric-energy \epem collider at SLAC. 
  Approximately $\nbb \times 10^6$ \BB pairs recorded at the \FourS
  resonance were used.

  We reconstruct $\Bz \to \KKKz$ decays by combining two oppositely-charged kaon
  candidates with a \Kz reconstructed as
  $\KS\to\pip\pim$ (\KKKspm)~\cite{Aubert:2004ta}, $\KS\to\piz\piz$ (\KKKszz), or $\KL$ (\KKKll). 
  Each $\KS\to\piz\piz$ candidate is formed from two
  $\piz\to\gamma\gamma$ candidates. 
  Each photon has 
$E_\gamma > 50 \mev$ and transverse shower shape 
  consistent with an electromagnetic shower. 
  Both \piz candidates satisfy 
$100 < m_{\gamma\gamma} <
  155 \mevcc$ and yield 
  an invariant mass $m_{\piz \piz}$ in the range $-20 < m_{\piz
    \piz} - m_{\KS} < 30 \mevcc$. 
  A \KL candidate is defined by an unassociated energy deposit in the electromagnetic
  calorimeter or an isolated signal in the Instrumented Flux Return~\cite{Aubert:2005ja}.

  For each fully reconstructed \Bz meson (\BCP), we use the remaining
  tracks in the event to reconstruct the decay vertex of the other \B
  meson (\Btag), and to identify its flavor \qtag~\cite{Aubert:2004zt}. 
  For each event we calculate
  the difference $\deltat \equiv t_{\CP} - t_{\mathrm{tag}}$ between the proper
  decay times of the \BCP\ and \Btag\ mesons, and its uncertainty $\sigma_{\deltat}$.

  We characterize \KKKspm and \KKKszz candidates using two kinematic
  variables: the beam-energy-substituted mass \mes and the energy
  difference \DeltaE~\cite{Aubert:2005ja}. 
  The signal region (SR) is defined as $\mes>5.26$ \gevcc,
  and $|\DeltaE|< 0.06$ \gev for \KKKspm, or $-0.120 < \DeltaE < 0.06$
  \gev for \KKKszz. 
  For \KKKll the SR is defined by $-0.01 < \DeltaE < 0.03 \gev$~\cite{Aubert:2005ja}, 
 and the missing
  momentum for the entire event is required to be consistent with the calculated
  \KL laboratory momentum.

  The main source of background is  
  continuum $e^+e^-\to \qqbar~ (q=u,d,s,c)$ events. 
We use event-shape variables to exploit the jet-like structure of these
  events in order to remove much of this background~\cite{Aubert:2005ja}.

  We perform an unbinned maximum likelihood fit to the selected \KKKz
  events using the likelihood
  function defined in Ref.~\cite{Aubert:2005ja}.
  The probability density function (PDF), ${\cal P}_i$, 
is given by
\begin{eqnarray}
 \lefteqn{{\cal P}_i \equiv  {\mathcal P}(\mes)\cdot {\mathcal P}(\DeltaE) \cdot {\mathcal
    P}_{\mathrm{Low}}} \\
  & &\cdot~ {\cal P}_{DP}(\mKK, \cos\theta_H, \deltat, \qtag) \otimes {\cal
    R}(\deltat, \sigma_{\deltat}), \nonumber
\end{eqnarray}
  where $i$ = (signal, continuum, \BB background), and
  ${\mathcal R}$ is the \deltat\ resolution function~\cite{Aubert:2004zt}. 
  For \KKKll, ${\mathcal P}(\mes)$ is not used. 
  ${\mathcal  P}_{\mathrm{Low}}$ is a 
PDF used only in the
  low-mass fit, which depends on the event-shape
  variables and, for \KKKll only, the missing momentum in the event~\cite{Aubert:2005ja}. 
  We characterize \Bz (\Bzb) events on the \DP in terms of
  \mKK and \cosH, the cosine of the helicity angle between the $\Kp$ ($\Km$) and the
  $\Kz$ ($\Kzb$) in the rest frame of the $\Kp\Km$ system.
  The \DP PDF for signal events is
  \begin{equation}
    {\mathcal P}_{DP} = d\Gamma \cdot \varepsilon(\mKK, \cosH) \cdot | J |,
  \end{equation}
  where $d\Gamma$ is the time- and flavor-dependent decay rate over the 
  \DP, $\varepsilon$ is the efficiency, and
  $J$ is the Jacobian of the transformation 
  to our choice of \DP coordinates.

  The time- and flavor-dependent decay rate is
  \begin{eqnarray}
\frac{d\Gamma}{d\deltat}
     \propto  \frac{e^{-|\deltat|/\tau}}{2\tau} &\times&  
    \Big[~ \left | {\cal A} \right |^2 + \left | \bar{ {\cal A} } \right |^2  \label{eq::dalitz_plot_rate}  \\ 
    &+&  ~  \qtag  ~2 \mathrm{Im} \left (\xi \bar{\cal A} {\cal A}^* \right ) \sin \deltamd \deltat \nonumber \\
    &-&  ~ \qtag \left (\left | {\cal A} \right |^2 - \left | \bar{ {\cal A} } \right |^2 \right ) \cos\deltamd\deltat ~\Big ], \nonumber
  \end{eqnarray}
  where $\tau$ and \deltamd are the lifetime and mixing frequency of the \Bz meson, respectively~\cite{PDG}. 
The parameter $\xi=\eta_{\CP} e^{-2i\beta}$, where $\beta =
\arg(-V_{cd} V_{cb}^*/V_{td} V_{tb}^*)$ and $V_{qq^\prime}$ are CKM
matrix elements~\cite{CKM}.
  The \CP eigenvalue $\eta_{\CP} = 1~(-1)$ for the \KS (\KL) mode. 
  We define the amplitude \AorAbar for \BzorBzbar decay as a sum of isobar amplitudes~\cite{PDG}, 
  \begin{eqnarray}
    \lefteqn{\AorAbar(\mKK,\cosH) = \sum \limits_r {\AorAbar}_r} \label{eq:A} \\
&&= \sum \limits_r c_r (1 \mp b_r) e^{i (\varphi_r \mp
  \delta_r)} \cdot  f_r(\mKK,\cosH),\nonumber
  \end{eqnarray}
  where 
the minus signs are associated with the $\overline{\cal A}$,
the parameters $c_r$ and $\varphi_r$ are the magnitude and phase of the
  amplitude of component $r$, and we allow for different isobar
  coefficients for $\Bz$ and $\Bzb$ decays through the asymmetry
  parameters $b_r$ and $\delta_r$.  

  Our isobar model includes resonant amplitudes $\phi$,
  $f_0$, $\chi_{c0}(1P)$, and 
  $X_0(1550)$~\cite{Aubert:2006nu, Garmash:2004wa}; 
  non-resonant terms; and 
  incoherent terms for \Bz decay to $\Dm \Kp$
  and $\Dsm \Kp$. 
For each resonant term, 
the function $f_r = F_r \times T_r \times Z_r$ describes the dynamical
  properties, where $F_r$ is the Blatt-Weisskopf centrifugal barrier factor for the
  resonance decay vertex~\cite{blatt}, $T_r$ is the resonant mass-lineshape, and
  $Z_r$ describes the angular distribution in the decay~\cite{Zemach:1963bc}.
 The barrier factor
  $F_r = 1/\sqrt{1+(Rq)^2}$~\cite{blatt} for the $\phi$, where $\vec{q}$ is the
  \Kp momentum in the $\phi$ rest frame and $R=1.5~{\gev}^{-1}$; 
  $F_r = 1$ for the scalar resonances. 
For $\phi$ decay $Z_r \sim \vec{q} \cdot \vec{p}$, where
  $\vec{p}$ is the momentum of the \Kz in the $\phi$
  rest frame, while $Z_r=1$ for the scalar decays.
  We describe the $\phi$, $X_0(1550)$, and
  $\chi_{c0}(1P)$ with relativistic Breit-Wigner lineshapes~\cite{PDG}.
  For the $\phi$ and $\chi_{c0}(1P)$ parameters we use average
  measurements~\cite{PDG}. For the $X_0(1550)$ resonance, we use
  parameters from our analysis of the $\Bp \to \KKK$ decay~\cite{Aubert:2006nu}.
  The $\fz$ resonance is described by a coupled-channel amplitude~\cite{Flatte:1976xu},
  with the parameter values of Ref.~\cite{Ablikim:2004wn}.

We include three non-resonant (NR) amplitudes 
  parameterized as $f_{\mathit{NR},k} =   \exp(-\alpha m^2_{k})$,
  where the parameter $\alpha = 0.14 \pm 0.01~c^4/\gev^2$ is taken from 
   measurements of $\Bp \to \KKK$ decays with larger signal samples~\cite{Garmash:2004wa,Aubert:2006nu}. 
  We include a complex isobar coefficient for each component $k = ( \Kp\Km, \Kp\Kz, \Km\Kz )$.

PDFs for \qqbar background in $\Bz \to \KKKs$
are modeled using
  events in the region $5.2 < \mes < 5.26 \gevcc$. The region
  $0.02 < \DeltaE < 0.04 \gev$ is used for \KKKll.
Simulated \BB events are used to define \BB background PDFs.
  We use two-dimensional histogram PDFs to model the \DP distributions for \qqbar and \BB backgrounds.

  We compute the \CP asymmetry parameters for component $r$ from the asymmetries in amplitude ($b_r$) 
  and phase ($\delta_r$) given in Eq.~(\ref{eq:A}).  
  The rate asymmetry is
  \begin{equation}
    \Acpr=\frac{|\bar{\cal A}_r|^2 - |{\cal A}_r|^2}{|\bar{\cal A}_r|^2 + |{\cal A}_r|^2} =\frac{-2b_r}{1+b_r^2},
    \label{eq:Acp}
  \end{equation}
  and $\betaeffr = \beta + \delta_r$ is the phase asymmetry.

  The selection criteria yield 3266 \KKKspm, 1611 \KKKszz, and 27513 \KKKll candidates 
which we fit to obtain 
the event yields, the isobar coefficients of the \DP model,
  and the \CP asymmetry parameters averaged over the \DP. The
  parameters $b_r$ and $\delta_r$ are constrained to be the same for
all model components, so in this case $\Acpr = \Acp$ and $\betaeffr = \betaeff$.
  We find $947 \pm 37$ \KKKspm, $144 \pm 17$ \KKKszz, and $770 \pm 71$ \KKKll signal 
  events. 
  Isobar coefficients and fractions are reported in
  Table~\ref{tab:isobars}, and \CP asymmetry results are
  summarized in Table~\ref{tab:cp}.
  The fraction ${\cal F}_r$ for resonance $r$ is computed as in Ref.~\cite{Aubert:2006nu}.
  Note that there is a $\pm \pi \rad$ ambiguity in the $\chi_{c0}(1P)\Kz$ phase.

\begin{figure}[hbt]
    \begin{center}
        \epsfig{file=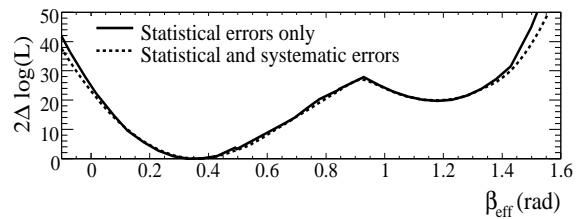,width=8cm,height=3cm}
      \caption{The change in
        twice the negative log likelihood as a function of \betaeff
        for the fit to the whole \DP.\label{fig:betascan}}
    \end{center}
\end{figure}

    \begin{table}[thb]
    \center
   \caption{The isobar amplitudes $c_r$, phases $\varphi_r$, and fractions ${\cal F}_r$ from the fit to the full \KKKz \DP. 
      The three NR components are combined for the fraction
      calculation. Errors are statistical only. Because of
      interference, $\sum {\cal F}_r\ne 100 \%$.}
    \begin{tabular}{|l|rrr|}
      \hline \hline
      Isobar Mode         &     Amplitude $c_r$     &       Phase $\varphi_r~(\rad)$    &       ${\cal F}_r$ (\%)\\
      \hline  
      $\phi\Kz$     & $  0.0085\pm 0.0010$          &       $ -0.016 \pm 0.234$    &       $12.5 \pm 1.3$            \\ 
      $\fz\Kz$      & $  0.622 \pm 0.046$           &       $ -0.14 \pm 0.14$      &       $40.2 \pm 9.6$            \\
      $X_0(1550)\Kz$  & $  0.114 \pm 0.018$           &       $ -0.47 \pm 0.20$      &       $4.1 \pm 1.3$           \\ 
      $(\Kp\Km)_{\mathit{NR}} \Kz$  & 1 (fixed)                     &       0 (fixed)              &                                 \\
      $(\Kp\Kz)_{\mathit{NR}} \Km$    & $  0.33 \pm 0.07$           &       $  1.95  \pm 0.27$     &       $112.0 \pm 14.9$          \\
      $(\Km\Kz)_{\mathit{NR}} \Kp$     & $  0.31 \pm 0.08$           &       $ -1.34  \pm 0.37$     &                                 \\
      \hline
      $\chi_{c0}(1P)\Kz$  & $  0.0306\pm 0.0049$    &     $ ^{\phantom{-}0.81}_{-2.33} \pm 0.54$    &       $3.0 \pm 1.2$           \\
    $\Dm\Kp$          & $  1.11\pm 0.17$                &                            &       $3.6 \pm 1.5$           \\
     $\Dsm\Kp$          & $  0.76\pm 0.14$               &                            &       $1.8 \pm 0.6$           \\
      \hline \hline
    \end{tabular}
     \label{tab:isobars}
  \end{table}
  In Fig.~\ref{fig:betascan}, we plot twice the change in the negative
  logarithm of the likelihood as a function of \betaeff.
  We find that 
  the \CP-conserving case of $\betaeff = 0$ is excluded at
  \CPsigma ($5.1\sigma$), including statistical and systematic errors (statistical errors only). 
 Also, the interference between \CP-even and \CP-odd
  amplitudes leads to the exclusion of the \betaeff solution near $\pi/2 - \beta$
  at \reflectionsigma ($4.6\sigma$).

  \begin{figure}[tbh]
    \begin{center}
\begin{tabular}{lr}
\epsfig{file=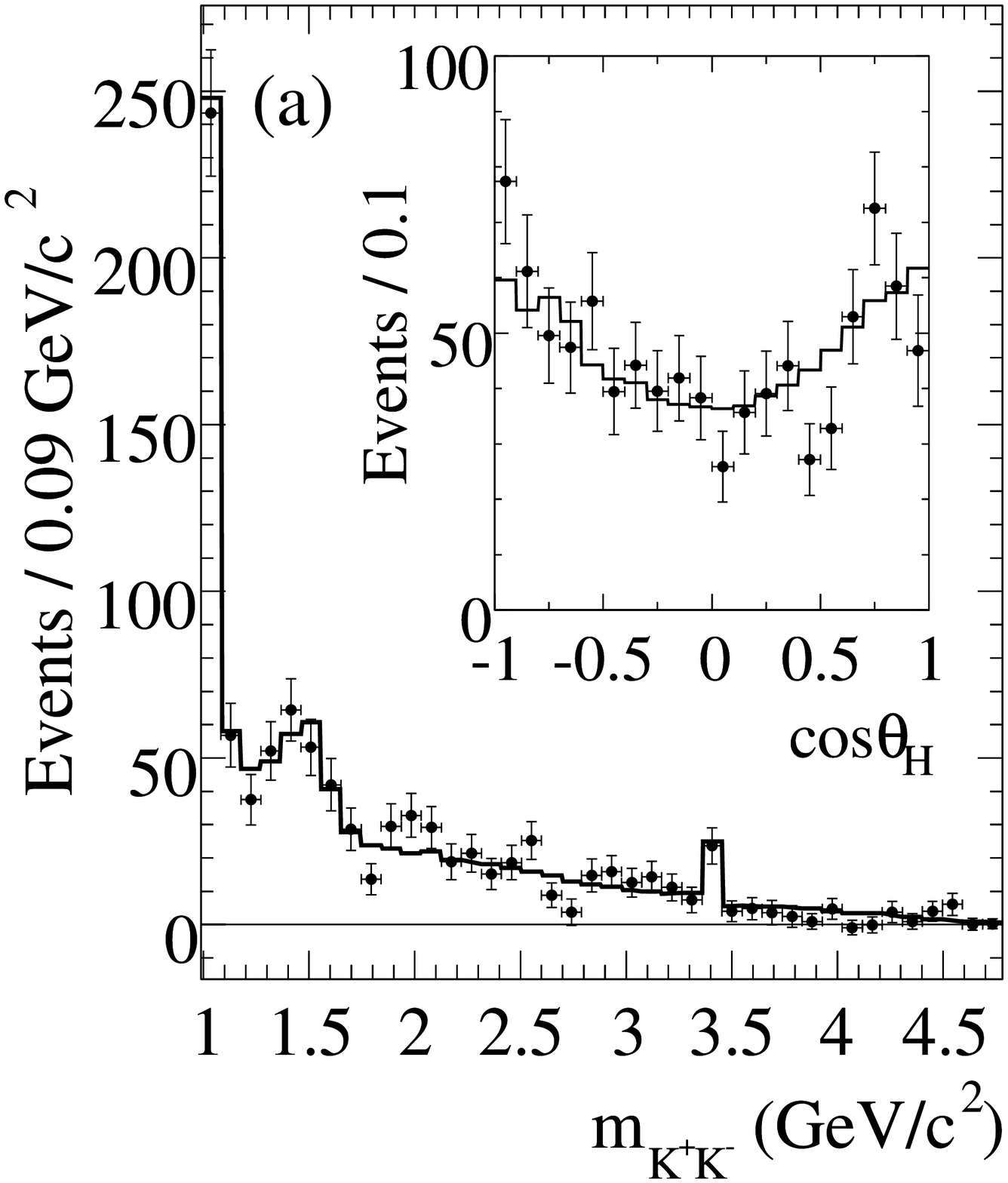,width=4.2cm,height=5.2cm}  &
\epsfig{file=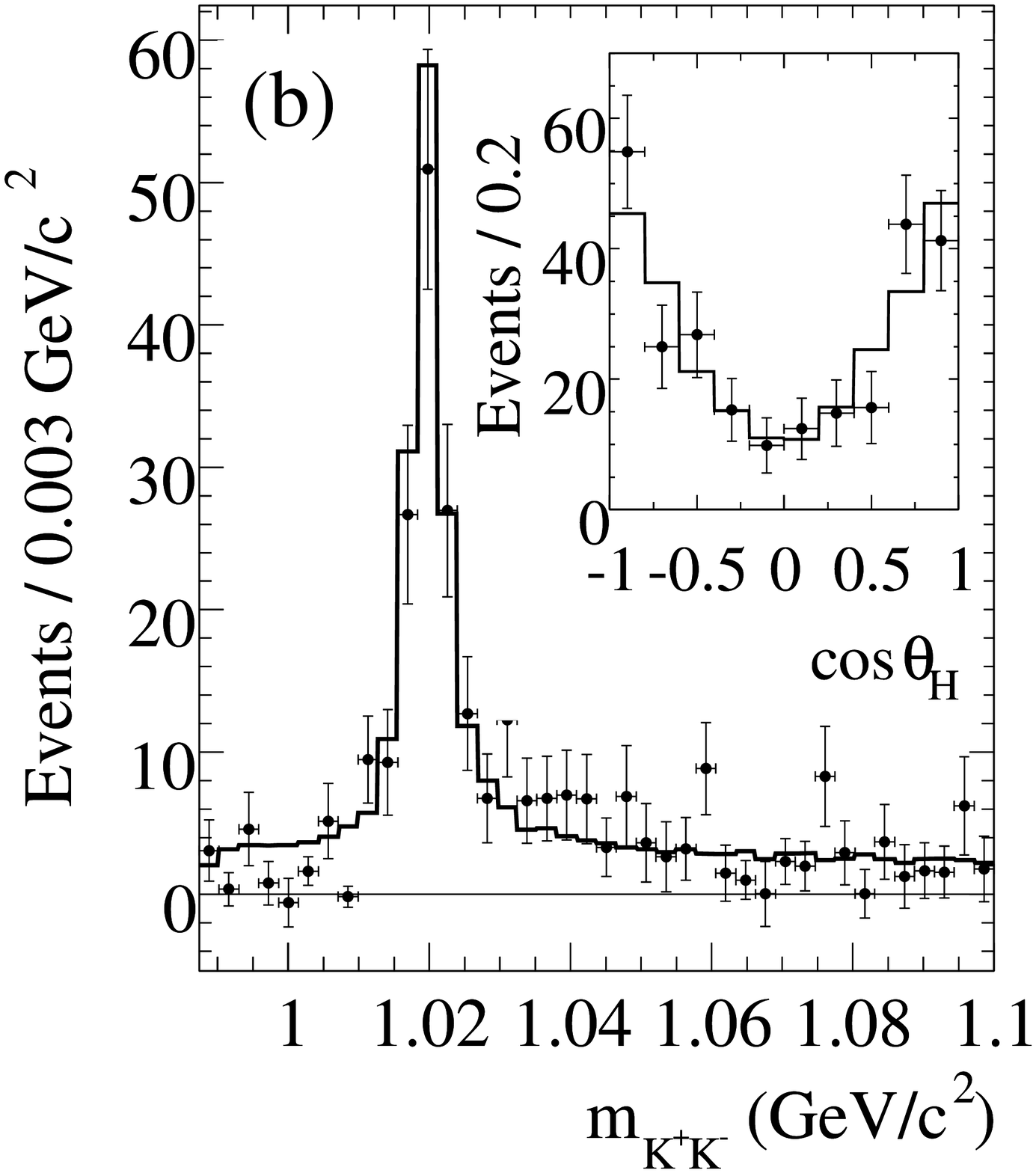,width=4.2cm,height=5.2cm}
\end{tabular}
\caption{The distributions of \mKK for
  signal-weighted~\cite{Pivk:2004ty} \KKKspm data in (a) the entire DP
  and (b) the low-mass region. Insets show distributions of
  \cosH. The histograms are projections of the fit function for
    the corresponding result.\label{fig:dalitz-splot}}
    \end{center}
  \end{figure}
  \begin{figure}[tbh]
    \begin{center}
 \epsfig{file=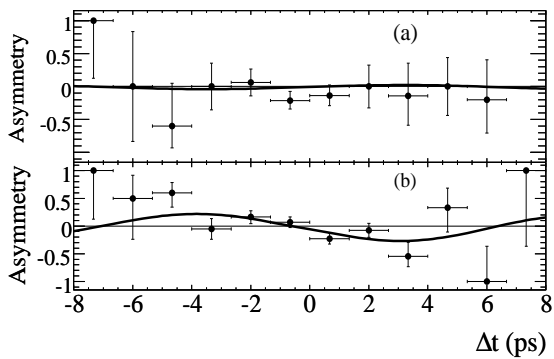,width=8cm,height=5cm}
 \caption{The raw asymmetry between 
   \Bz- and \Bzb-tagged signal-weighted~\cite{Pivk:2004ty}
   events for \KKKspm, in (a) the low-mass region
   and (b) the high-mass region. The curves are
   projections of the corresponding fit results.
\label{fig:deltat}}
    \end{center}
  \end{figure}
  We also measure \CP asymmetry parameters for events with $\mKK<1.1$ \gevcc.
  In this region, we find 1359 \KKKspm, 348 \KKKszz, 
  and 7481 \KKKll candidates.
  The fit yields $282 \pm 20$, $37 \pm 9$ and $266 \pm 36$ signal events, 
  respectively.
  The most significant contributions in this region are from
  $\phi\Kz$ and $\fz\Kz$ decays, with a smaller contribution from
  the low-mass tail from non-resonant decays.
  In this fit we vary the amplitude asymmetries $b_r$ and
  $\delta_r$ for 
  the $\phi$ and $\fz$, while the other components are fixed to the SM
  expectations of
  $\betaeff=0.370~\rad$ and $\Acp=0$~\cite{hfag}. We also vary the isobar coefficient for the $\phi$,
  while fixing the others to the results from the whole DP fit.
  There are two solutions with likelihood difference of only $\Delta \log L = 0.1$.
  Solution~(1) is consistent with the SM, while in
  Solution~(2) \betaeff for the \fz differs significantly from the SM
  value (Table~\ref{tab:cp}). The solutions also differ significantly
  in the values of the $\phi$ isobar coefficient.
  There is also a mathematical ambiguity of $\pm\pi \rad$ on
  \betaeff for the $\phi$, with a corresponding change of $\pm\pi \rad$ in the
  solution for $\varphi_{\phi}$. This 
  ambiguity is present for both solutions.
The fit correlation between the $\phi$ and \fz in $\delta_r$ is $0.71$~\cite{epaps}.

Finally, we perform a fit to extract the average \CP asymmetry parameters
in the high-mass region. In the 2384 \KKKspm, 1406 \KKKszz, and 20032
\KKKll selected events with $\mKK>1.1~\gevcc$, we find signal yields of
$673 \pm 31$, $87 \pm 14$ and $462 \pm 56$ events, respectively; the
\CP asymmetry results are shown in Table~\ref{tab:cp}. 
We find that
for this fit the \CP-conserving case of $\betaeff = 0$ is excluded at
$5.1\sigma$, including statistical and systematic errors.

Figure \ref{fig:dalitz-splot} shows distributions of the \DP
variables \mKK and \cosH obtained using the 
method described in \cite{Pivk:2004ty}. Figure \ref{fig:deltat} shows 
the \deltat-dependent asymmetry between \Bz- and \Bzb-tagged events.
  \begin{table}[htb]
    \center
    \caption{The \CP-asymmetries for
      $\Bz\to\KKKz$ for the entire DP, in the high-mass region, and for $\phi\Kz$ and $\fz\Kz$
in the low-mass region. The first errors are statistical and the
second are systematic. The solutions (1) and (2) from the low-mass fit are discussed in the text.\label{tab:cp}}
     \begin{tabular}{|l|c|c|}
      \hline\hline
                   & \Acp   &     $\betaeff (\rad)$       \\                   
      \hline
      Whole DP    &  $\wholeDPacp$&    $\wholeDPbeta$   \\
\hline
      High-mass    &  $-0.054 \pm 0.102 \pm 0.060$ &  $0.436 \pm 0.087\,^{+0.055}_{-0.031}$   \\
      \hline
      (1) $\phi\Kz$ &  $-0.08 \pm 0.18 \pm 0.04$ & $0.11 \pm 0.14 \pm 0.06$    \\
      (1) $f_0\Kz$  &  $\phantom{-}0.41 \pm 0.23 \pm 0.07$   & $0.14 \pm 0.15 \pm 0.05$  \\
\hline
      (2) $\phi\Kz$ &  $-0.11 \pm 0.18 $ & $0.10 \pm 0.13$    \\
      (2) $f_0\Kz$  &  $-0.20 \pm 0.31$   & $3.09 \pm 0.19$  \\
      \hline\hline
    \end{tabular}
 \end{table}

Systematic errors on the \CP-asymmetry parameters are listed in Table~\ref{tab:systematics}.
The fit bias uncertainty includes effects of detector resolution and possible 
  correlations among the fit variables determined from full-detector 
simulations.
We also account for uncertainties due to the isobar model:
experimental precision of resonance parameter values; alternate $X_0(1550)$
parameter values~\cite{Garmash:2004wa}; and, in the low- and high-mass fits,
the statistical uncertainties on the isobar coefficients determined in
the fit to the whole DP. 
Other uncertainties common to many \babar\ time-dependent analyses, including
those due to fixed PDF parameters, and possible \CP asymmetries in the
\BB background are also taken into
account~\cite{Aubert:2005ja,ref:tagint}.  Uncertainties due to fixed
PDF parameters are evaluated by shifting the fixed parameters and
refitting the data.
As a cross-check, we perform the analysis using \KKKspm alone and find
results consistent with those in Table~\ref{tab:cp}.
  \begin{table}[htb]
    \center
    \caption{A summary of the systematic errors on the \CP asymmetry parameter values.}
    \begin{tabular}{|l|rr|rr|rrrr|}
      \hline \hline
      Source    &       \multicolumn{2}{c|}{Whole DP} &  \multicolumn{2}{c|}{High-mass} &  \multicolumn{2}{c}{$\phi\Kz$} & \multicolumn{2}{c|}{$f_0\Kz$} \\
      & $\Acp$  &  \betaeff        &  $\Acp$    &  \betaeff &   $\Acp$   &  \betaeff    &   $\Acp$   &  \betaeff     \\ \hline 
      \hline
      Fit Bias  & 0.003   &   0.001    &  0.014   &  0.008  &  0.03  & 0.06 & 0.06 & 0.03    \\
 Isobar model   & 0.004   &   0.009    &  0.025   &  $^{+0.051}_{-0.024}$  &  0.00  & 0.01 & 0.01 & 0.03  \\
      Other     & 0.052   &   0.024    &  0.053   &  0.018  &  0.02  & 0.01 & 0.03 & 0.02 \\
      \hline
      Total     & 0.053   &   0.026    &  0.060   &  $^{+0.055}_{-0.031}$  &  0.04  & 0.06 & 0.07 & 0.05  \\

      \hline \hline
    \end{tabular}
    \label{tab:systematics}
  \end{table}

  In summary, in a sample of $\nbb \times 10^6$ \BB meson pairs we
  simultaneously analyze the
  \DP distribution and measure the 
  time-dependent \CP asymmetries for $\Bz\to\KKKz$ decays.
The values of \betaeff and \Acp are consistent with the SM expectations of 
  $\beta \simeq 0.370~\rad,~\Acp \simeq 0$~\cite{hfag}. The
  signficance of \CP violation is \CPsigma, and we reject the solution near
  $\pi/2 -\beta$ at \reflectionsigma.
We also
  measure \CP asymmetries for the decays $\Bz \to \phiKz$ and $\Bz \to
  \fz \Kz$, where we find \betaeff lower than the SM expectation by about $2\sigma$.
  The \CP parameters in the high-mass region are compatible with SM
  expectations, and we observe \CP violation at the
  level of $5.1\sigma$.
\par
  
We are grateful for the excellent luminosity and machine conditions
provided by our \pep2\ colleagues, 
and for the substantial dedicated effort from
the computing organizations that support \babar.
The collaborating institutions wish to thank 
SLAC for its support and kind hospitality. 
This work is supported by
DOE
and NSF (USA),
NSERC (Canada),
CEA and
CNRS-IN2P3
(France),
BMBF and DFG
(Germany),
INFN (Italy),
FOM (The Netherlands),
NFR (Norway),
MIST (Russia),
MEC (Spain), and
STFC (United Kingdom). 
Individuals have received support from the
Marie Curie EIF (European Union) and
the A.~P.~Sloan Foundation.


\begin{thebibliography}{99}
\bibitem{conjugate}
  Use of charge conjugate reactions is implied throughout.
\end{thebibliography}
\end{document}